\documentclass[lettersize,journal]{IEEEtran}
\usepackage{diagbox}
\usepackage{amsmath}    
\usepackage{algorithm}
\usepackage{algorithmic}
\usepackage{multirow}

\usepackage{svg}
\usepackage{multicol}
\usepackage{caption}
\usepackage{float}
\usepackage{chemformula}
\usepackage{color}
\usepackage{pifont}

\usepackage{mdframed}
\usepackage{algorithm}
\usepackage{algorithmic}
\usepackage[version=4]{mhchem}
\usepackage{tikz}
\tikzset{
main node/.style={inner sep=0,outer sep=0},
label node/.style={inner sep=0,outer ysep=.2em,outer xsep=.4em,font=\scriptsize,overlay},
strike out/.style={shorten <=-.2em,shorten >=-.5em,overlay}
}

\PassOptionsToPackage{bookmarks={false}}{hyperref}

   \usepackage{graphicx}
   \graphicspath{{./Figures/}}
   \DeclareGraphicsExtensions{.pdf,.jpeg,.png,.fig, .emf}
   \usepackage{subfigure}
    \usepackage{epstopdf}
    \usepackage{epsfig}
    \usepackage{subfigure}
    \usepackage{epstopdf}
    \usepackage{epsfig}
    \usepackage{cancel}
    \usepackage{amsmath}

\usepackage{RobStd}
\HideNotes
\usepackage{makecell}
\usepackage{booktabs}
\usepackage{multirow}
\usepackage{subcaption}

\usepackage{tabularx}

\newif\ifsubmit
\submitfalse

\definecolor{comments}{rgb}{0.1, 0.66, 0.1}
\ifsubmit
    \newcommand{\can}[1]{}
\else
    \newcommand{\can}[1]{[{\color{blue}CL: #1}]}
\fi

\ifsubmit
    \newcommand{\xu}[1]{}
\else
    \newcommand{\xu}[1]{[{\color{purple} XU: #1}]}
\fi

\newcommand\eat[1]{}

\begin{document}

\title{Efficient and Variation-Tolerant Tree Inference via Compact and Scalable Mapping in Analog CAMs}

\author{
\small
Zhicheng Xu$^{1,4}$, Wenqing Shi$^{1,4}$, Pengyu Ren$^{2}$, Chao Li$^{3}$, Bo Wen$^{1,4}$, Ruibin Mao$^{1,4}$, Xunzhao Yin$^{3}$, Xiaobo Sharon Hu$^{2}$, Can Li$^{1,4,*}$\\
$^1$Department of Electrical and Computer Engineering, The University of Hong Kong, Hong Kong SAR, China\\
$^2$Department of Computer Science and Engineering, University of Notre Dame, Notre Dame, IN, USA\\
$^3$College of Information Science and Electronic Engineering, Zhejiang University, Hangzhou, China\\
$^4$Center for Advanced Semiconductors and Integrated Circuits, The University of Hong Kong, Hong Kong SAR, China\\
$^*$Corresponding authors, email: canl@hku.hk 
}

\maketitle
\begin{abstract}
Tree-based machine learning methods are widely adopted across various domains due to their inherent interpretability and, specifically on tabular data, their state-of-the-art accuracy. Despite their advantages, these methods face efficiency challenges on conventional architectures due to irregular memory access patterns during tree traversal operations. Prior work utilizes analog Content Addressable Memories (CAM) to solve the problem, but its application is limited by poor scalability due to inefficient mapping, 
where increasing tree depth requires exponentially more analog CAM column resources. Their accuracy is also  degraded by device non-idealities such as programming variations and soft decision boundaries. 

To address these challenges, we propose ACTIVAR, a compact variation-tolerant tree implementation in dense analog CAM array via software-hardware codesign. 
On the software side, we introduce an analog-CAM-aware compact tree model that employs hierarchical subtrees with shared features to maximize hardware utilization. 
This dense mapping introduces redundancy, improving both model accuracy and hardware defect tolerance.
Additionally, we develop Variation-Enhanced Training, which utilizes a fuzzy overlap boundary strategy to build models resilient to analog noise. On the hardware side, we design a FeFET-based analog CAM cell with a driver to sharpen decision thresholds and implement a pipelined multi-core architecture to optimize throughput and energy consumption. 
Across 13 classification and regression datasets, ACTIVAR maintains near-software-level accuracy while reducing hardware overhead by up to 99\% compared with prior analog-CAM-based soft-tree designs. Under realistic hardware noise conditions, ACTIVAR limits accuracy degradation to just 3.7\% on the MNIST dataset, whereas baseline models suffer a catastrophic 49.6\% drop. Furthermore, the ACTIVAR architecture achieves energy efficiency gains ranging from $10^4$ to $10^6$ times over commercial general-purpose processors, establishing it as a highly efficient solution for high-capacity machine learning at the edge.
\end{abstract}

\section{Introduction}
\label{sec:intro}

Tree-based machine learning methods, including decision trees (DT) and their corresponding ensemble techniques like random forests\cite{breiman2001random} and eXtreme Gradient Boosting (XGBoost)\cite{chen2016xgboost}, are widely adopted in diverse domains due to their interpretability. 
Moreover, specifically for tabular data, they remain the state-of-the-art (SOTA) in terms of accuracy performance. 
Compared with multilayer perceptrons (MLPs), tree-based models offer three key advantages. Primarily, they are more effective at capturing irregular decision boundaries through recursive partitioning\cite{grinsztajn2022tree}. Secondly, they are less sensitive to non-informative features, which is particularly important in high-dimensional settings where feature selection is difficult\cite{grinsztajn2022tree}. Moreover, their transparent decision process makes them well-suited for high-consequence domains such as healthcare, autonomous driving, and legal systems\cite{floridi2021establishing}. These advantages have established tree-based ensemble methods as the SOTA paradigm for tabular data processing across both scholarly research and industrial implementations.

However, to enhance predictive performance, these methodologies typically incorporate numerous decision trees, necessitating extensive branch-split operations during tree traversal that generate irregular memory access patterns\cite{xie2021tahoe}. This computational paradigm introduces substantial latency and energy inefficiencies when implemented on conventional von Neumann architectures, where the physical separation of compute and memory units incurs significant energy expenditure and temporal overhead for data transfer operations between these disparate units. The memory-processor dichotomy thus presents a critical bottleneck for tree-based ensemble methods, constraining their deployment in real-time high-throughput applications\cite{summers2020fast,chen2021catboost} and resource-constrained edge computing environments. 

Emerging computing architectures have demonstrated the potential to overcome this bottleneck. Recent work by Li et al. \cite{li2020analog} and Yin et al. \cite{yin2020fecam} leverages analog content-addressable memory (analog CAM) with resistive RAM (RRAM) and ferroelectric field-effect transistors (FeFETs) to implement in-memory decision tree acceleration. Thanks to the high parallelism and in-situ comparison capability of analog CAMs, tree-based methods can be accelerated to more than $120\times$ speedup over NVIDIA V100 GPU\cite{pedretti2024x}. However, existing analog CAM-based accelerator designs confront two fundamental limitations. 
\textit{First}, they suffer from limited scalability due to inefficient sparse mapping. 

As the maximum tree depth increases, the required analog CAM array size grows exponentially, while most analog CAM cells remain unutilized\cite{pedretti2021tree, wen2025trustworthy}. 
\textit{Second}, the attainable model accuracy is severely compromised by non-ideal characteristics in emerging memory. Programming variation in analog conductance states induces imprecise threshold control, while inherent physical limitations, such as non-ideal subthreshold swing, create probabilistic matching behavior \cite{yin2024deep}.  This probabilistic nature establishes intermediate "soft boundary" regions where feature matching outcomes are indeterminate, falling between fully matched and mismatched states, thereby degrading the overall prediction fidelity. 

To address the challenges, we propose ACTIVAR, a hardware-aware framework for compact and variation-tolerant tree inference in analog CAMs. Our design addresses the structural sparsity in traditional decision trees, enabling deeper tree deployment under the same analog CAM resource budget through improved cell utilization.
By incorporating a Fuzzy Overlap Boundary (FOB) during training, the model maintains software-level predictive accuracy despite non-idealities experimentally observed in physical devices. System-level evaluations show that our pipelined multi-core architecture not only bridges the accuracy gap between software and hardware but also outperforms general-purpose processors in energy efficiency by up to six orders of magnitude.

The contributions of this work are itemized as follows:
\begin{itemize}
\item 
We propose a novel hardware-aware compact tree model that resolves the structural mismatch between irregular tree structure and regular analog CAM arrays. By enforcing feature-sharing at identical tree depths, our model achieves up to 100\% analog CAM cell utilization within subtrees, overcoming the exponential hardware resource growth associated with deep decision trees.
\item We introduce a Variation-Enhanced Training (VET) algorithm utilizing a "Fuzzy Overlap Boundary" (FOB) strategy. This method replicates borderline samples during training to ensure decision robustness against stochastic device variations and input voltage noise inherent in analog CAM implementations.
\item We design the ACTIVAR accelerator, featuring a FeFET-based analog CAM cell with integrated 4T drivers to sharpen ``soft" decision boundaries. The architecture employs a pipelined multi-core design and a Network-on-Chip (NoC) interconnect, enabling high-throughput parallel tree traversal and selective core activation for extreme energy efficiency.
\item We evaluate ACTIVAR across 13 diverse classification and regression datasets. Experimental results demonstrate that ACTIVAR maintains near-software accuracy while reducing hardware resource consumption by up to 99\% compared to state-of-the-art analog CAM designs. Furthermore, our framework achieves 
$10^4\times$ to $10^6\times$ energy efficiency relative to conventional high-performance CPU architectures.
\end{itemize}

\section{Background}

\subsection{Tree-based Models Acceleration}
\begin{figure}
    \centering
    \includegraphics[width=\linewidth]{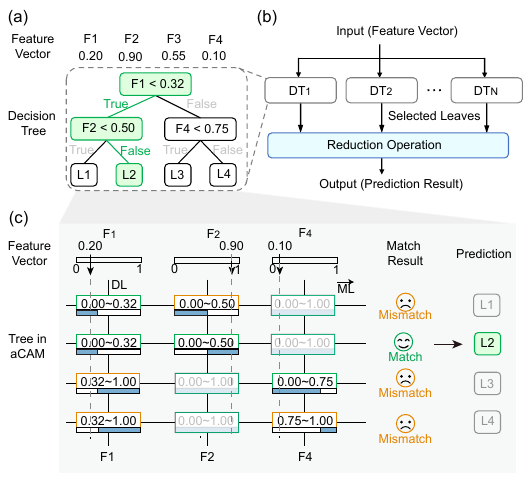}
    \caption{Tree-based model acceleration in analog CAM. (a) Structure of a Decision Tree (DT). (b) Structure of a tree ensemble model. (c) Mapping of the DT in (a) onto an analog CAM array.}
    \label{fig:dt}
    \vspace{-1em}
\end{figure}

Decision Tree (DT) is a fundamental machine learning method valued for its simplicity, effectiveness on small datasets, and high interpretability, which allows domain experts to understand and verify the model logic easily. As illustrated in \autoref{fig:dt}(a), a typical DT structure consists of internal nodes that compare an input feature against a predefined threshold. Based on the comparison result, the input is routed to the appropriate successor node in the next layer. This process continues until a leaf node is reached, which provides the final prediction. 
These models can be further enhanced through ensemble techniques. For example, \autoref{fig:dt}(b) depicts a Random Forest (RF) model\cite{breiman2001random}, a DT ensemble in which all trees perform inference in parallel. Due to their ability to capture irregular patterns and model non-smooth functions in tabular data, tree-based methods continue to outperform deep learning models in specific domains, particularly those involving structured data with heterogeneous features\cite{grinsztajn2022tree}.

However, the acceleration of tree-based models suffers from irregular and unpredictable memory access patterns in hardware, creating a major bottleneck when computing tree-based models on GPUs \cite{he2022booster}. To address this challenge, Li et al. proposed acceleration using analog CAM\cite{li2020analog}. As shown in \autoref{fig:dt}(c), each root-to-leaf path in the DT is mapped to one row in the CAM array, and the input feature vectors are fed into the array to be processed by all rows in parallel. 
This work has been extended to accelerate Random Forest models \cite{pedretti2021tree,pedretti2024x} and, more recently, Deep Random Forest architectures \cite{yin2024deep}. Furthermore, recent advancements have explored utilizing 3D NAND-Flash to achieve higher parallelism for tree-based inference \cite{zhong2024nand}.
Despite these advances, significant challenges remain for real-world applications. The analog matching method is vulnerable to device and voltage variations, compromising reliability. Furthermore, the required CAM array size grows exponentially when the tree depth increases, leading to an extreme sparsity that most CAM cells remain unutilized. For example, more than 95\% of the cells are not used during the implementation of a tree model with depth of 8, which reveals substantial potential for hardware efficiency gains. 

\subsection{Analog Content Addressable Memories}
Recently, various CAM designs have been proposed\cite{mahendra2017self, li2020analog, yin2020fecam, yin2022ferroelectric, liu2024analog,xu2024ferex} for parallel search, using both standard CMOS and diverse emerging non-volatile devices. These designs enable the in situ parallel search of stored data (organized in rows) against input queries. Among these approaches, analog CAM \cite{yin2020fecam, li2020analog, pedretti2021tree, yin2024deep} stands out due to its higher memory density, higher energy efficiency, and improved flexibility compared to conventional digital implementations.
Unlike the digital CAM, such as Ternary CAM (TCAM), which stores discrete state values, analog CAM stores a continuous range defined by upper and lower bounds. These bounds are encoded via the configurable devices in analog CAM cells like memristors \cite{li2020analog} or FeFETs \cite{yin2020fecam}.

\begin{figure}
    \centering
    \includegraphics[width=\linewidth]{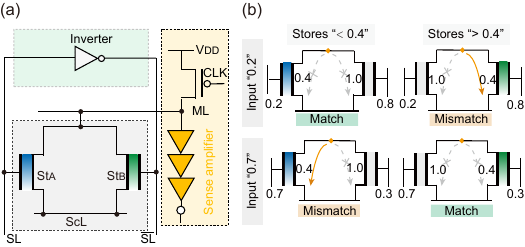}
    \caption{(a) The 2FeFET-based analog CAM design. (b) Examples of the way of an analog CAM cell to store the lower bound and upper bound.}
    \label{fig:cam}
    \vspace{-0.5em}
\end{figure}

\autoref{fig:cam}(a) illustrates the 2FeFET analog CAM design \cite{yin2020fecam}. Before a search operation, the match line (ML) is precharged. During search, an input voltage $V_{SL}$ is applied, and its complement (via an on-cell inverter) is fed to the second FeFET. The left (blue) FeFET sets the upper bound, while the right (green) FeFET, responsive to the inverted signal, sets the lower bound. As shown in \autoref{fig:cam}(b), if $V_{SL}$ lies outside the stored range (i.e., above the upper bound or below the lower bound), one of the FeFETs turns on and discharges the ML, signaling a mismatch. If $V_{SL}$ falls within the bounds, both FeFETs remain off, the ML stays charged, and a match is registered. This configuration enables efficient range-based comparisons using analog thresholds.
However, these decision boundaries are not perfectly sharp; instead, they exhibit soft transitions due to analog device physics. Consequently, inputs near the threshold regions produce ambiguous match/mismatch outcomes. Furthermore, device-to-device variations can shift the effective bounds, introducing additional reliability challenges.

\subsection{FeFET Device}
\begin{figure}
    \centering
    \includegraphics[width=\linewidth]{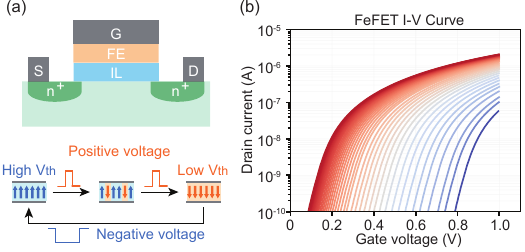}
    \caption{(a) The structure of the FeFET device, illustrating its variable threshold voltage ($V_{th}$) resulting from partial polarization of the ferroelectric layer. (b) The I-V curve of the FeFET device, showcasing its electrical characteristics.}
    \label{fig:back_fefet}
    \vspace{-1em}
\end{figure}

FeFETs have garnered significant research interest in the in-memory computing community due to their exceptional performance characteristics: high ON/OFF ratios exceeding $10^6$, fast write/read speeds, high endurance, ultra-low energy consumption, high density, and CMOS compatibility \cite{li2024high,muller2012ferroelectricity,khan2020future,ni2018critical}. 
\autoref{fig:back_fefet}(a) illustrates the FeFET structure, which incorporates \ce{HfO2} as the ferroelectric dielectric layer within a conventional MOSFET gate stack\cite{dunkel2017fefet}. This integration introduces a distinctive operational characteristic, the polarization behavior of the ferroelectric layer, which differentiates FeFETs from standard MOSFETs. FeFETs utilize the polarization direction in the ferroelectric layer to store memory states. When a positive gate voltage pulse is applied, an n-type FeFET accumulates electrons in the nucleation-dominated channel, causing domains in the ferroelectric layer to probabilistically switch to the positive state, which reduces the device's threshold voltage ($V_{th}$). Conversely, applying a negative voltage increases the FeFET's $V_{th}$.

By varying pulse width parameters, the threshold voltage ($V_{th}$) can be continuously modulated, enabling analog state representation (see \autoref{fig:back_fefet}(b)). Leveraging this analog characteristic, previous work\cite{yang2024energy,yin2024deep,yin2020fecam} has demonstrated high-density in-memory computing implementations based on FeFETs. However, these analog states come with significant challenges. Device-to-device variations make precise programming difficult. Accurately distinguishing between closely spaced analog states remains problematic\cite{mulaosmanovic2018mimicking, ni2020impact, garg2021impact}. These limitations constrain the practical deployment of analog FeFET-based computing systems despite their theoretical advantages.

\section{ACTIVAR SOFTWARE DESIGN}
This section introduces ACTIVAR's analog CAM-aware tree model design, which focuses on scalability and resilience against device variation. We first present a compact tree model that addresses the low CAM cell utilization problem in traditional designs, enabling the implementation of deep, powerful trees on analog CAM. We then introduce Variation-Enhanced Training to mitigate accuracy loss caused by device variation.
\subsection{Analog CAM-Aware Compact Tree Model}
\label{sec:compact}
\begin{figure}
    \centering
    \includegraphics[width=\linewidth]{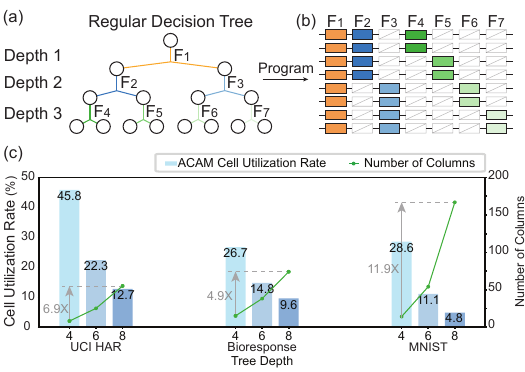}
    \vspace{-2em}
    \caption{Sparsity problem in regular decision trees implemented on analog CAM accelerators: (a) feature selection pattern mismatch with analog CAM structure, (b) exponential increase in required columns and declining utilization with tree depth, making deep powerful tree models on analog CAM practically infeasible.}
    
    \label{fig:depthscale}
\end{figure}
\begin{figure}
    \centering
    \includegraphics[width=\linewidth]{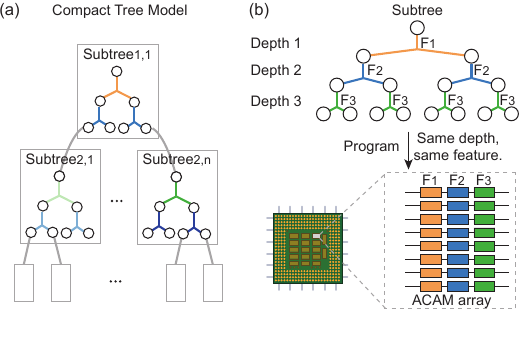}
    \vspace{-2em}
    \caption{Proposed analog CAM-aware compact tree. (a) Hierarchical structure of compact tree with multiple subtrees, each programmable on a separate analog CAM array. (b) Structure of a subtree where all nodes at the same depth share the same feature, enabling dense programming to analog CAM.}
    \vspace{-1em}
    \label{fig:ODT}
\end{figure}

\begin{algorithm}[t]

\caption{Analog CAM-aware Compact Tree Training}

\label{alg:compacttree}
\begin{mdframed}[linecolor=black,linewidth=1pt]

\textbf{Input:} $\mathcal{D}$: Dataset; $C$: analog CAM cols; $N$: Subtree count; $L$: Max levels

\textbf{Output:} $T$: Decision tree

1:  Let $T \gets \emptyset$, $\mathcal{S} \gets \emptyset$, queue $Q \gets {(\mathcal{D}, 0)}$;

2:  \textbf{while} $Q$ not empty and level $l < L$ \textbf{do}

3:  \quad Dequeue $(\mathcal{D}_{\text{cur}}, l)$ from $Q$;

4:  \quad Let $S$ $\gets$ subtree with depth $C$ at level $l$;

5:  \quad \textbf{for} each depth $d = 1$ to $C$ \textbf{do}

6:  \quad \quad Greedily select feature $f_d$ for all nodes at depth $d$;

7:  \quad \textbf{end for}

8:  \quad Add $S$ to $\mathcal{S}$;

9: \quad ${\mathcal{D}_{\text{leaves}}} \gets \text{Partition (}\mathcal{D}_{\text{cur}}, S, Strategy); $ 

10: \quad Enqueue $\{(\mathcal{D}_{\text{leaf}}, l+1) \mid \mathcal{D}_{\text{leaf}} \in {\mathcal{D}_{\text{leaves}}}\}$ to $Q$;

11: \textbf{end while}

12: \textbf{while} $|\mathcal{S}| > N$ \textbf{do}

13: \quad Prune lowest-gain subtree from $\mathcal{S}$;

14: \textbf{end while}

15: Build $T$ from $\mathcal{S}$;

16: \textbf{return} $T$;
\end{mdframed}

\end{algorithm}

\autoref{fig:depthscale}(a) illustrates the sparsity problem that arises when mapping conventional decision trees onto analog CAM accelerators. In depth-growing trees, each node independently selects its own feature to minimize local impurity, resulting in irregular and unstructured feature-access patterns. However, analog CAM arrays operate on a fixed structure, where each feature corresponds to a dedicated column. This mismatch forces the accelerator to allocate separate analog CAM columns for every distinct feature used across the tree. As the tree becomes deeper, the diversity of feature selections increases rapidly, causing an exponential growth in required columns and a sharp decline in cell utilization. For example, as shown in \autoref{fig:depthscale}(b), increasing the depth from 4 to 8 enlarges the required column count by more than $11\times$, while utilization drops to only $4.8\%$. \textbf{This structural inefficiency makes deploying deep, high-capacity tree models on analog CAM hardware impractical.}

To overcome this scalability bottleneck, we reorganize a deep tree into a hierarchy of subtrees that fit the available analog CAM resources, as shown in \autoref{fig:ODT}a. 

Within each subtree, all nodes at the same depth share the same feature, which matches the fixed-column structure of analog CAM arrays and enables 100\% cell utilization (\autoref{fig:ODT}b). Unlike conventional oblivious trees, this feature-sharing constraint is applied only locally within each subtree. Different subtrees are programmed on separate analog CAM arrays, enabling the use of distinct features across subtrees to maintain the overall model capacity and accuracy. In addition, this design also simplifies training, since subtrees at the same level can be trained separately once the data have been partitioned by the previous level.

The training method is presented in  \autoref{alg:compacttree}. Our approach constructs the compact tree model according to the available analog CAM resources, adhering to column count $C$ , maximum subtree count $N$ (which scales linearly with available array count) and max subtree levels $L$. The training process proceeds level by level, with each subtree's depth constrained by the number of columns in the analog CAM array. At each level, we employ a breadth-first strategy when selecting features, ensuring that nodes at the same depth within a subtree utilize identical features. After constructing a subtree, the corresponding dataset is partitioned according to the leaf nodes, creating refined datasets for subsequent subtrees. To address the hardware non-idealities, the partitioning step will be enhanced with a variation-aware strategy, which is detailed in \autoref{sec:VAE}. 
This hierarchical process continues iteratively until either the maximum level 
$L$ is reached or no further partitioning is needed.

When the resulting model exceeds the hardware constraint of $N$ subtrees, we implement a pruning strategy that removes subtrees with the lowest information gain, ensuring both accuracy and efficient hardware resource utilization.

\subsection{Variation-Enhanced Training }
\label{sec:VAE}
\begin{figure}
    \centering
    \includegraphics[width=\linewidth]{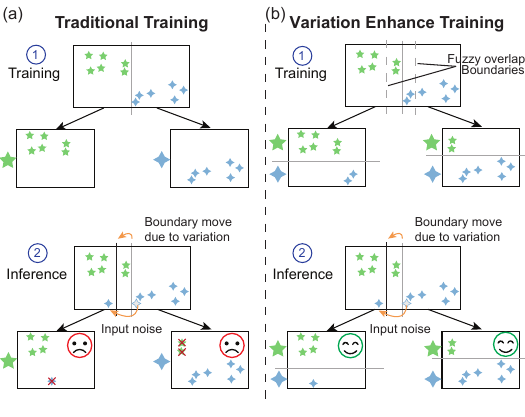}
    \caption{Comparison between traditional training and our Variation-Enhanced Training. (a) Traditional training: device variation or input noise causes misrouting of borderline data points to untrained paths. (b) VET: fuzzy overlap boundaries ensure that data near thresholds is processed by both subtrees, mitigating the impact of device variation and input noise.}
    \label{fig:VAR}
\end{figure}

\begin{algorithm}[t]

\caption{Variation-Enhanced Training}

\label{alg:VET}
\begin{mdframed}[linecolor=black,linewidth=1pt]
\textbf{INPUT:} Dataset $\mathcal{D}$, Fuzzy overlap boundary $\delta$ \\
\textbf{OUTPUT:} Tree $T$ 

1: \textbf{function} BuildTree($node$, $\mathcal{S}$, $\delta$)

2: \quad \textbf{if} stopping criteria met \textbf{then}

3: \quad \quad \textbf{return} node as leaf with majority class;

4:  \quad  \textbf{end if}

5: \quad Let $gain_{best} \gets -\infty$;

6: \quad \textbf{for} each feature $f$, threshold $thr$ in $\mathcal{S}$ \textbf{d}o

7: \quad \quad Let $\mathcal{S}_{gain} \gets \{ x \mid x_f \notin [thr - \delta, thr + \delta] \}$;

8: \quad \quad Let $g \gets$ gain($\mathcal{S}_{gain}$, impurity);

9: \quad \quad \textbf{if} $g > gain_{best}$ \textbf{then}

10: \quad \quad \quad Update $gain_{best}$, $thr_{best}$, $f_{best} $;

11: \quad \quad \quad Let $\mathcal{S}_{left} \gets \{ x \mid x_f \leq thr + \delta \}$;

12: \quad \quad \quad Let $\mathcal{S}_{right} \gets \{ x \mid x_f > thr - \delta \}$;

13: \quad \quad \textbf{end if}

14: \quad \textbf{end for}

15: \quad Let $node.feature \gets f_{best}$, $node.threshold \gets thr_{best}$;

16: \quad Call BuildTree($node.left$, $\mathcal{S}_{left}$, $\delta$);

17: \quad Call BuildTree($node.right$, $\mathcal{S}_{right}$, $\delta$);

18: \textbf{end function}

19: \textbf{return } BuildTree($node$, $\mathcal{S}$, $\delta$) \textcolor{blue}{// Main function}
\end{mdframed}

\end{algorithm}

Another significant improvement in our tree model is its variation-tolerant design, which mitigates the substantial accuracy drop when the decision tree is implemented on analog CAM hardware. This design stems from our observation of how device variation in analog CAM causes significant accuracy degradation. 

In analog CAM hardware, the threshold of each tree node is encoded in an emerging memory device. Device variation can therefore perturb the effective threshold during inference. As a result, samples near a decision boundary can be routed to the wrong branch during inference. In a conventional tree, that branch has never seen those samples during training, so even a small threshold shift can cause a large accuracy drop (See \autoref{fig:VAR}a).

To enable the model to handle misrouted data, particularly those near decision boundaries, we propose a novel method called Variation-Enhanced Training (VET). Unlike traditional training, which splits the dataset using a precise threshold and trains subsequent nodes with the resulting subdatasets, our method introduces a Fuzzy Overlap Boundary (FOB) on both sides of the threshold. When splitting the dataset for the next node, data within this sensitive boundary is sent to both the left and right subtrees. This ensures that both subtrees are trained to process these sensitive data points, mitigating the impact of device variation during inference on analog CAM hardware. Additionally, this method not only effectively addresses device variation but also enhances robustness to input noise in the dataset, improving overall model performance. 

The detailed implementation is shown in \autoref{alg:VET}. It is important to note that this fuzzy partitioning strategy is not limited to individual node training; it can be used in partitioning logic in \autoref{alg:compacttree} (Line 9). By applying the FOB when distributing data to subsequent subtrees, we ensure that boundary samples are duplicated into the training sets of multiple potential hardware cores. Consequently, even if circuit non-ideality causes a borderline sample to be misrouted to an unintended subtree, that subtree is already trained to process the input correctly, neutralizing the routing error. 
However, this duplication is used only for recursive training, not for split selection. Information gain is computed using only the samples outside the fuzzy boundary, where the routing decision is reliable. This approach ensures that feature selection is driven by data points with clear classification signals rather than borderline cases, consequently enabling the tree to learn decisions based on the most stable features under variation/noise. This distinction is reflected in \autoref{alg:VET}: $S_{\mathrm{gain}}$ is used for gain computation, while $S_{\mathrm{left}}$ and $S_{\mathrm{right}}$ include the duplicated boundary samples for recursive training. In practice, the boundary width $\delta$ is determined by the expected device variation and input noise, and its effect is analyzed in Section~\ref{sec:robust}.

\section{ACTIVAR Hardware design}

We present the hardware components of ACTIVAR in this section. The section is organized as follows: first, we discuss the FeFET analog CAM cell with sharper boundaries; next, we describe the analog CAM array for executing a single subtree; and finally, we outline the full chip integrating an analog CAM-aware compact tree.
\subsection{Analog CAM Cell with Sharper Boundary}
\label{sec:shaper_acam_cell}
\begin{figure}
    \centering
    \includegraphics[width=\linewidth]{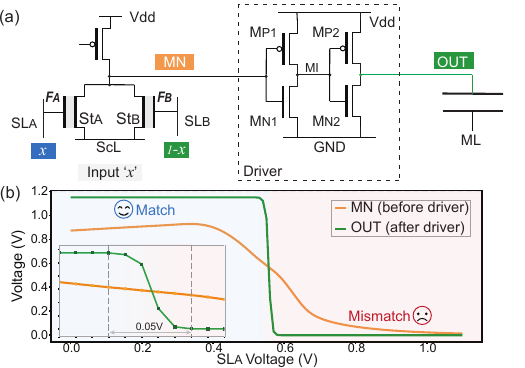}
    \caption{{\textbf{(a)}} The proposed FeFET-based analog CAM cell with Mitigated Soft Boundary (MSB), featuring a FeFET pair module, 4T Schmitt trigger, and a capacitor. {\textbf{(b)}} Input-output voltage characteristics of the Driver. {\textbf{(c)}} Voltage response at the Match Node (MN) and OUT terminals as the search line voltage changes, illustrating the sharp transition boundary.}
    \label{fig:camcell}
\end{figure}

In addition to device variation, another significant disparity between analog CAM hardware and software implementations is the soft boundary in analog CAM cell caused by subthreshold swing, which can reduce accuracy. To address the soft boundary issue, we develop an analog CAM cell design with sharper boundaries, integrating a 4T driver that interfaces the FeFET pair module—comprising two parallel-connected FeFETs ($F_A$ and $F_B$) controlled via Search Lines ($SL_A $and $SL_B$)—with a capacitor, as illustrated in \autoref{fig:camcell}(a).

The FeFET pair module enables analog comparison at each tree node, where high voltage at the match node (MN) indicates a match and otherwise a mismatch. To implement the comparison, $F_A$ stores the upper threshold voltage, and $F_B$ stores the lower threshold. Given a FeFET memory window of $[0,V]$ and an input $x\in[0,1]$ , the comparison operates as follows: \ding{182} $xV$ is applied to $F_A$, resulting in a mismatch if $xV > V_{th,A}$ (i.e., when $x > V_{th,A}/V$); \ding{183} $(1-x)V$ is applied to $F_B$, yielding a mismatch if $(1-x)V > V_{th,B}$ (i.e., when $x < 1-V_{th,B}/V$). 

The output of the FeFET pair is sharpened by the 4T driver using two inverters, which pull intermediate voltages to either VDD(High voltage, $V_H$) or GND(Low voltage, $V_L$), effectively eliminating the soft boundary region. The effectiveness of our design is demonstrated in Fig.~\ref{fig:camcell}(b), which shows the output voltage $V_{out}$ response to search line voltage variations while maintaining constant FeFET threshold voltages. The results confirm that our design reduces the soft boundary to just 4.2\% of the full memory window (0.05V out of 1.2V range), representing a significant improvement in boundary sharpness compared to conventional analog CAM implementations.

\subsection{Analog CAM Array Design}
\begin{figure}
    \centering
    \includegraphics[width=\linewidth]{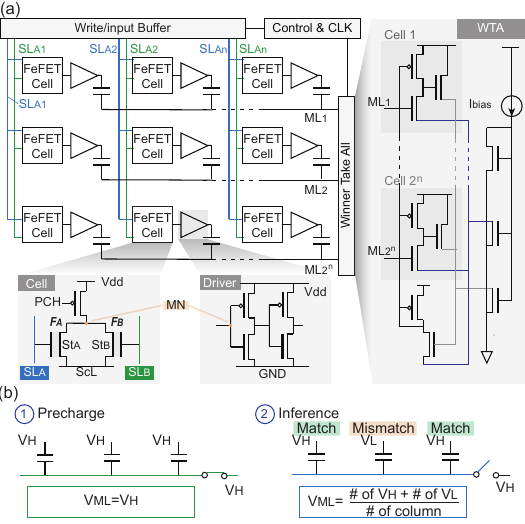}
    \caption{\textbf{(a)} The proposed analog CAM structure. Matching outcomes are registered as ML voltages, and the best result is then selected by a WTA. \textbf{(b)} Two-phase operation: (1) Precharge phase, initializing all MLs to $V_H$, and (2) Inference phase, computing match levels through capacitive coupling.}
    \label{fig:circuit}
\end{figure}
Building on the sharpened analog CAM cell, the analog CAM array is designed to execute a subtree of the compact tree efficiently, as illustrated in \autoref{fig:circuit}(a). 
The array consists of $2^n \times n$ (rows × columns) analog CAM cells, a write/input buffer, and a multi-input voltage-based winner-take-all circuit \cite{soleimani2011voltage}. 
To ensure the high parallel tree inference, analog CAM cells in the same column share a common search input ($SL_A$ and $SL_B$, while cells in the same row have their capacitors interconnected to produce match level results on the match line (ML) through capacitor coupling\cite{ma2022capcam}.  

The inference process in the analog CAM array, depicted in \autoref{fig:circuit}(b), proceeds in three steps:
\begin{enumerate}

\item \textbf{Precharge Phase}: A precharge (PCH) signal sets each cell’s OUT node to high voltage ($V_H$) and connects the ML to a voltage source at ($V_H$), establishing the initial match line voltage.

\item \textbf{Inference Phase}: The ML is disconnected and left floating. Search signals are applied to the search lines, triggering comparisons with the stored thresholds in each analog CAM cell. The results produce either high (match) or low (mismatch) voltage levels at individual cell outputs. For each mismatch in a row, the ML voltage decreases by $\frac{V_H-V_L}{\#\text{ of columns}}$, resulting in a final ML voltage that directly correlates with the match quality:
\begin{equation}
V_{ML} = V_H - \frac{\#\text{ of mismatches}}{\#\text{ of columns}} \times (V_H - V_L)
\end{equation}
\item \textbf{Decision Phase}: The winner-take-all circuit identifies the row with the highest ML voltage, corresponding to the path with the most matches, thus determining the subtree output.
\end{enumerate}

\subsection{Analog CAM-aware Compact Tree Architecture}

\begin{figure}
    \centering
    \includegraphics[width=\linewidth]{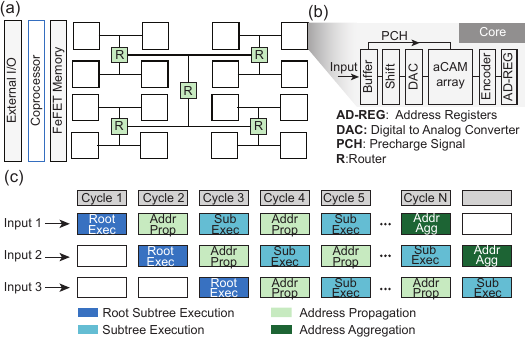}
    \caption{ACTIVAR multi-core accelerator architecture. (a) System overview with co-processor, FeFET memory, and cores connected via H-tree NoC. (b) Core architecture with analog CAM array, buffer, shift register, DAC, and address register. (c) Four-stage pipeline execution: root subtree execution, address propagation, subtree execution, and address aggregation.}
    \label{fig:arch}
\end{figure}

To scale the analog CAM array's capabilities for implementing the complete ACTIVAR tree structure across multiple subtrees, we propose the ACTIVAR multi-core accelerator architecture, illustrated in \autoref{fig:arch}(a). This architecture integrates FeFET memory with multiple specialized processing cores interconnected through an H-tree Network-on-Chip (NoC) topology utilizing distributed routers (R). A central co-processor orchestrates the entire system, managing data coordination and external I/O interface operations.

Each processing core incorporates an analog CAM array coupled with input buffers, shift registers for sequential data operations, and a Digital-to-Analog Converter (DAC) to handle the analog signal requirements of the analog CAM cells, as detailed in \autoref{fig:arch}(b). Upon completion of the analog CAM array's decision-making phase, the resulting match line signals undergo processing through a digital encoder that transforms the winner-take-all output into a digital address, subsequently stored in the Address Register (AD-REG). This address serves as the key for accessing stored decision outcomes and guiding subsequent tree traversal operations.

The execution methodology for the analog CAM-aware compact tree implements a pipelined, hierarchical approach designed to optimize both performance metrics and power efficiency. Each core assumes responsibility for executing a designated subtree within the comprehensive tree architecture. The pipeline operates through multiple stages with coordinated timing to minimize processing latency while maximizing system throughput, effectively accommodating tree structures with varying subtree configurations (\autoref{fig:arch}(c)).

The execution sequence proceeds through the following operational phases:
\begin{enumerate}
\item \textbf{Root Subtree Execution (Cycle 1)}: The co-processor initiates execution of the root subtree in the designated core. The analog CAM array performs parallel comparisons, and the winner-take-all circuit generates an address output, which is stored in the AD-REG.
\item \textbf{Address Propagation (Cycle 2)}: The co-processor retrieves the address from the AD-REG and uses it to determine the next subtree to activate. This address is routed through the NoC to the appropriate core.
\item \textbf{Subtree Execution (Cycles 3 to N-1)}: The selected core executes the next subtree, repeating the analog CAM array matching process. This step is selectively activated to avoid redundant computations in irrelevant branches, reducing power consumption. For a tree with depth ( D ), this stage may repeat for multiple subtrees (up to ( D-1 ) iterations), with each subtree execution producing an address that guides the traversal to the next subtree.
\item \textbf{Address Aggregation (Cycle N)}: Once all required subtrees in the traversal path are processed, the co-processor aggregates the collected addresses to form a composite address. This address accesses the FeFET memory to retrieve the final prediction result corresponding to the leaf node.
\end{enumerate}


\section{Hardware Evaluation}
\subsection{Model Calibration and Simulation Framework}

To evaluate the energy consumption and delay characteristics of our hardware design, we established a comprehensive Spectre simulation framework anchored by experimental calibration data. We calibrated a Preisach FeFET model\cite{ni2018circuit} using experimental measurements from real FeFET analog CAM cell fabricated in GlobalFoundries' 28-nm gate-first high-K metal gate CMOS process. The fabricated FeFET devices feature a gate stack composed of poly-crystalline Si/TiN (2 nm)/doped $HfO_2$ (8 nm)/$SiO_2$ (1 nm)/p-Si.

The experimental validation focused on characterizing individual FeFET pair operation, as illustrated in \autoref{fig:camcell}(a). During these measurements, we configured the $F_B$ transistor to the high-threshold voltage state ($V_{th}$ = 1.2 V) while varying the threshold voltage states of the $F_A$ transistor. These experimental results, shown in \autoref{fig:Functionality}(a), provided the foundation for calibrating our simulation models, with the calibrated model validation presented in \autoref{fig:Functionality}(b).

\subsection{Evaluation for Array Energy Consumption and Delay}

We evaluated the energy consumption and operational delay of our analog content-addressable memory system using the calibrated FeFET models in conjunction with 45-nm Predictive Technology Model (PTM) models for MOSFETs\cite{nimo2007}. 
The simulation framework included both the analog CAM array structures and peripheral circuits, with particular focus on the WTA circuit. FeFET parasitic capacitances were set to 0.05 fF based on established literature values\cite{yin2024deep}. Array configurations from $8\times 3$ to $256 \times 8$ elements were tested to assess scalability performance.

\begin{figure}
    \centering
    \includegraphics[width=\linewidth]{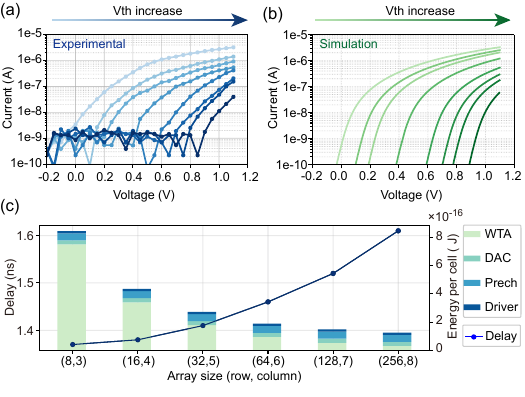}

    \caption{ \textbf{(a)} Experimental I-V curves of FeFET analog CAM cells with different $V_{th}$ states.  \textbf{(b)} Simulation validation using calibrated Preisach FeFET models.
\textbf{(c)} Array scalability analysis showing energy consumption breakdown and inference delay for different array sizes. The stacked bar chart displays energy components: WTA (Winner-Take-All), DAC (Digital-to-Analog Converter), Prech (Precharge), and Driver in analog CAM cell. }
    
    \label{fig:Functionality}

\end{figure}

System delay was evaluated under worst-case conditions, specifically when the minimum voltage gap exists between adjacent match line values $V_{ML}$. We define inference delay as the complete duration from CAM cell precharge initiation through winner identification by the WTA circuit. As demonstrated in \autoref{fig:Functionality}(c), the system delay exhibits a gradual increase with array size, rising from 1.37 ns for the smallest configuration to 1.62 ns for the largest 256×8 array. This moderate delay progression is primarily attributable to the increased decision-making time required by the WTA circuit as the array scale expands.

Inference energy is governed by the interplay between the crossbar array and the sensing module. The crossbar array imposes a constant computational overhead of 0.095 fJ/search/cell, independent of array size. This fixed cost aggregates contributions from cell precharge (0.050 fJ), search line charging (0.029 fJ), and driver operations (0.016 fJ). On the other hand, the energy of the sensing part (WTA circuit) changes with the array size. In small arrays, the WTA circuit consumes most of the total energy. However, as the number of elements increases, the cost of the WTA circuit is shared across more cells. As a result, the energy efficiency per search per cell improves by $6.7 \times$, dropping from 0.847 fJ/search/cell in the smallest design to 0.126 fJ/search/cell in the largest array. This trend shows that our analog CAM design becomes much more energy-efficient in larger applications.

\section{Software Evaluation}

In this section, we first evaluate the performance of the compact tree model across multiple datasets, and then discuss the influence of fuzzy overlap boundary and min samples split on the performance of the proposed compact tree model with variation-enhanced training. 

\subsection{Software Simulation Setup}
\begin{table}[t]

\centering
\caption{\centering DATASETS FOR CLASSIFICATION \\
($n$: FEATURE SIZE, $K$: NUMBER OF CLASSES)}
\label{tab:ClassificationDataset}
\begin{tabular}{|c|cccc|}
\hline
\textbf{Dataset} & $n$ & $k$ & \textbf{Instances} & \textbf{Description} \\
\hline
California\cite{california_housing_openml} & 8 & 2 & 20634 & Housing Price\\
Wine\cite{wine_109} & 13 & 3 & 178 & Wine Recognition \\
Breast Cancer\cite{breast_cancer_wisconsin_(diagnostic)_17} & 30 & 2 & 569 & Cancer Diagnosis \\
UCI HAR\cite{human_activity_recognition_using_smartphones_240}& 561 & 6 & 10299 & Activity Recognition \\
Fashion MNIST\cite{xiao2017fashion} & 784 & 10 & 70000 & Fashion Classification \\
MNIST\cite{lecun1998mnist} & 784 & 10 & 70000 & Digit Recognition \\
\hline
\end{tabular}
\vspace{-1em}
\end{table}

\begin{table}[t]
\centering

\caption{DATASETS FOR REGRESSION \\
($n$: FEATURE SIZE)}
\label{tab:RegressionDataset}
\begin{tabular}{|c|ccc|}
\hline
\textbf{Dataset} & $n$  & \textbf{Instances} & \textbf{Description} \\
\hline

California Housing\cite{pace1997sparse} & 8  & 20640 & Housing Price Regression \\
Abalone\cite{abalone_1} & 8  & 4177 & Age Prediction \\ Diabetes\cite{efron2004least} & 10 & 442 & Diabetes Diagnosis \\
Wine (White)\cite{wine_quality_186} & 11  & 4898 & Wine Quality Prediction \\
Wine (Red)\cite{wine_quality_186} & 11  & 1599 & Wine Quality Prediction \\
Appliances Energy\cite{appliances_energy_prediction_374} & 28  & 19735 & Energy Consumption \\
Student Performance\cite{student_performance_320} & 30  & 649 & Academic Performance \\
\hline
\end{tabular}
\vspace{-1em}
\end{table}
\textbf{Dataset Selection and Metrics:}
The evaluation encompasses diverse benchmarks spanning multiple domains. As detailed in \autoref{tab:ClassificationDataset} and \autoref{tab:RegressionDataset}, we selected six classification datasets and seven regression datasets, with feature dimensions ranging from 8 to 784 to cover various complexity levels.

For performance evaluation, classification tasks utilize prediction accuracy, while regression tasks employ normalized Root Mean Squared Error (RMSE). To enable meaningful cross-dataset comparisons, RMSE is calculated on target values min-max scaled to the [0, 1] range.

\textbf{Data Preprocessing:} 
Preprocessing protocols are standardized while accommodating dataset-specific needs. Categorical features are converted to numerical representations via label encoding. To manage computational complexity in high-dimensional datasets (e.g., UCI HAR, Fashion MNIST), the values of each continuous feature are discretized into 50 bins. Finally, all datasets undergo min-max scaling to ensure input consistency.

\textbf{Model Configuration:} The experimental design employs a two-level cascading compact tree structure with architecture parameters determined by training dataset size. To balance computational efficiency and model complexity, an '8+8' configuration (depth-8 root and subtrees) is applied to datasets with over 20,000 training samples, while a shallower '6+6' structure is used for smaller datasets to prevent overfitting.

\textbf{Baseline Configuration:} 
Standard decision tree classifiers and regressors from Scikit-learn serve as baselines. For fair comparison, all baseline trees are restricted to the exact same total depth as their compact tree counterparts.

\subsection{Evaluation for Compact Tree Model}
\begin{figure}
    \centering
    \includegraphics[width=\linewidth]{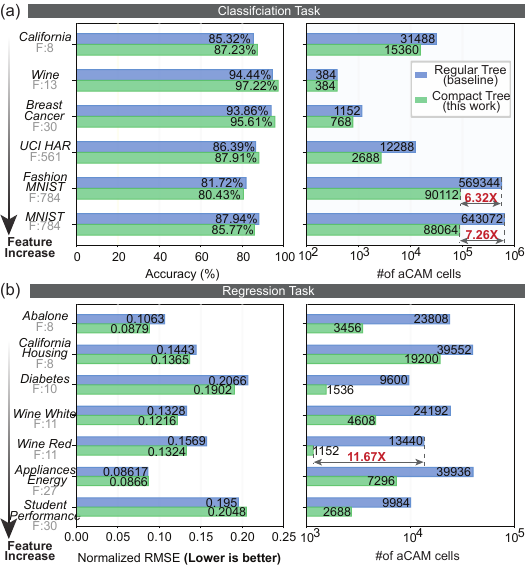}
    \caption{\textbf{(a)} Classification accuracy and analog CAM cell utilization for regular versus compact decision tree classifiers. \textbf{(b)} Normalized RMSE and analog CAM cell utilization for regular versus compact decision tree regressors. The compact design maintains comparable performance while significantly reducing hardware requirements across both task types.}
    \label{fig:model_accuracy}
\end{figure}
\begin{figure*}
    \centering
\includegraphics[width=0.95\linewidth]{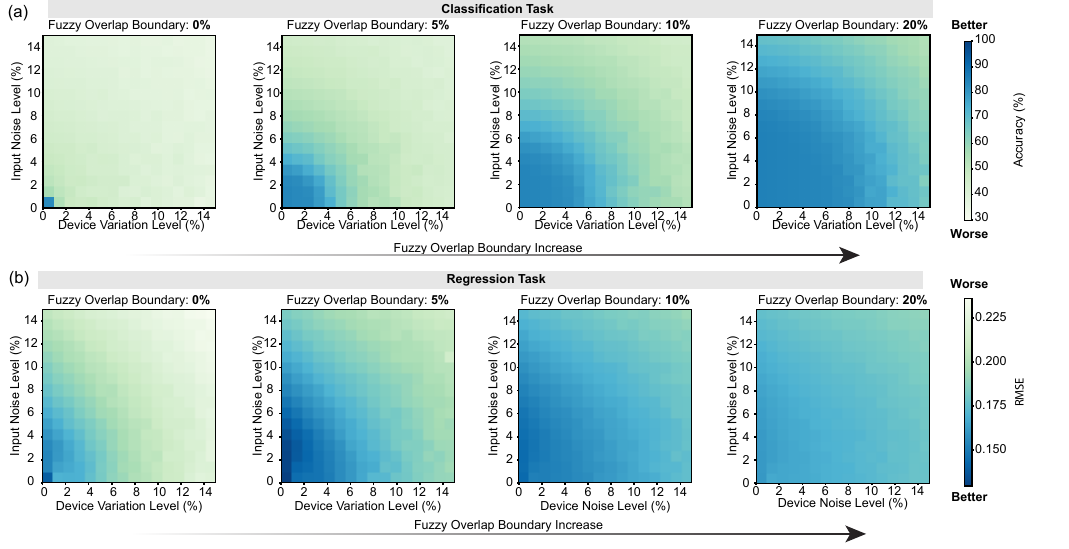}
    \vspace{-1em}
    \caption{\textbf{Impact of fuzzy overlap boundary on model robustness under hardware non-idealities.} Heatmaps showing \textbf{(a)} classification accuracy on MNIST and \textbf{(b)} RMSE on California Housing dataset across varying device variation and input noise for different FOB configurations. Darker blue indicates better performance.}
    \vspace{-2em}
    \label{fig:Evariation}
\end{figure*}
To evaluate the proposed training algorithm, we compared compact tree classifiers and regressors (configured with zero FOB) against Scikit-learn baseline decision trees. As shown in \autoref{fig:model_accuracy}, compact models achieve predictive performance comparable to baselines across diverse datasets while achieving a reduction of up to 91.4\% in hardware resource consumption.

Specifically, compact trees outperform regular trees on low-dimensional datasets for both classification and regression tasks. 
For instance, on the Student Performance regression task, the compact model yields lower RMSE despite using fewer resources. This performance gain stems from the architecture’s structural constraint: requiring all nodes at the same depth to share identical splitting features serves as a strong regularization technique, mitigating overfitting in limited feature spaces.

While a slight reduction in accuracy is observed on high-dimensional datasets, this is offset by a substantial decrease in analog CAM cell utilization, often by several orders of magnitude (right panels of \autoref{fig:model_accuracy}). This significant hardware efficiency provides sufficient margin to improve accuracy through deeper tree structures or ensemble methods (e.g., Random Forests) while maintaining low resource costs.

\subsection{Evaluate the Impact of Variation-Enhanced Training}

\label{sec:robust}

\begin{figure}
    \centering
    \includegraphics[width=\linewidth]{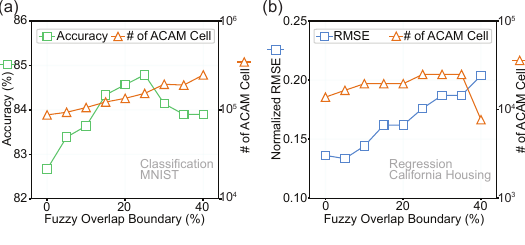}
    \caption{Impact of variation enhanced training on model prediction performance and analog CAM resource consumption across different fuzzy overlap boundary widths for \textbf{(a)} classification and \textbf{(b)} regression tasks (Lower RMSE means better prediction performance). }
    \label{fig:var_sweep}
    \vspace{-2em}
\end{figure}
To evaluate the efficacy of variation-enhanced training, we introduced device threshold variations and input noise (quantified as percentages of the feature range) to the MNIST and California Housing datasets.

\autoref{fig:Evariation}(a) demonstrates that our proposed training method enhances robustness against hardware non-idealities in classification tasks. Baseline models without variation-enhanced training (FOB = 0\%) are highly sensitive; classification accuracy on MNIST suffers a precipitous drop of nearly 31.7\% (from 82.4\% to 50.7\%) under just 1\% device variation and input voltage noise. Increasing the Fuzzy Overlap Boundary (FOB) width provides systematic resilience; for instance, an FOB of 20\% limits accuracy degradation to under 4\% even with 6\% device variation and input noise. Similar robustness improvements are observed in regression tasks, as shown in \autoref{fig:Evariation}(b).

This resilience, however, introduces a trade-off regarding prediction performance and hardware resource consumption (See \autoref{fig:var_sweep}(a) for classification task, \autoref{fig:var_sweep}(b) for regression). Small FOB widths act as a data augmentation strategy: by replicating samples within the fuzzy regions to both child nodes, the effective training set size increases, allowing the construction of deeper, more complex trees that capture finer decision boundaries. This results in slightly improved software accuracy but higher analog CAM consumption. Conversely, excessively wide boundaries reverse this trend. As the fuzzy region expands, fewer samples fall into the "distinct" non-overlapping intervals required for calculating split entropy. This lack of samples leads to premature termination of tree growth. Consequently, the final model becomes shallower and consumes fewer analog CAM cells, but suffers from underfitting and degraded prediction accuracy.

\subsection{Minimum Samples for Split}
\begin{figure}[h]
    \centering
    \includegraphics[width=\linewidth]{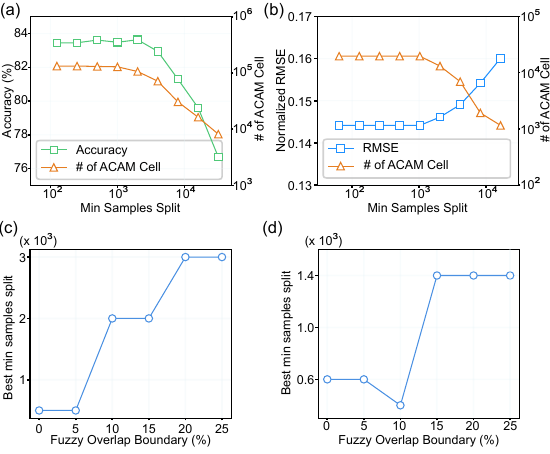}
    \caption{MSS parameter optimization for compact tree models. Trade-off between model prediction performance and analog CAM consumption versus MSS for \textbf{(a)} classification and \textbf{(b)} regression tasks. Best MSS values under different fuzzy overlap boundary settings for \textbf{(c)} classification and \textbf{(d)} regression tasks.}
    \vspace{-2em}
    \label{fig:mss}
\end{figure}

Variation-enhanced training improves robustness but leads to an increase in the tree size due to sample replication in fuzzy regions. To counteract this, Minimum Samples for Split (MSS) acts as a pruning parameter, balancing resource efficiency against predictive capability.

To illustrate the accuracy-consumption trade-off, we employed an "8+8" compact tree on MNIST with a 10\% FOB (\autoref{fig:mss}(a)). Increasing the root-level MSS up to 4000 samples reduces hardware requirements without compromising accuracy. This shifts the computational burden to the second-level subtrees, allowing the root to act as a coarse router. However, pushing MSS beyond this limit degrades performance. An overly pruned root passes excessive raw samples to the subtrees, exceeding their fitting capacity and causing underfitting. Similar trade-off in the regression task can be found in \autoref{fig:mss}(b).

Furthermore, we observed a positive correlation between the optimal MSS and the Fuzzy Overlap Boundary (FOB) width (\autoref{fig:mss}(c) and (d)). This relationship exists because the sample replication from a wider FOB increases the sample count at each node, allowing the model to accommodate a higher MSS threshold.

\section{Application and benchmarking}
To evaluate the proposed ACTIVAR design under realistic hardware conditions, we employ a simulation framework that incorporates device variation, input voltage noise, and soft decision boundaries. Based on experimental FeFET data\cite{soliman2020ultra}, we model a threshold ($\sigma_{th}$) of 54 mV within a 1.2 V memory window (approx. 4.5\%). Additionally, input voltage noise and decision softness are set to 4.5\% and 5\%, respectively. The following analysis compares ACTIVAR against standard Decision Trees (CART), and prior analog-CAM-based state-of-the-art implementations (soft tree \cite{wen2025trustworthy}) across classification and regression benchmarks with single-tree and ensemble methods, reporting the worst-case accuracy from 30 Monte Carlo simulations.

\begin{figure*}
    \centering
    \includegraphics[width=0.95\linewidth]{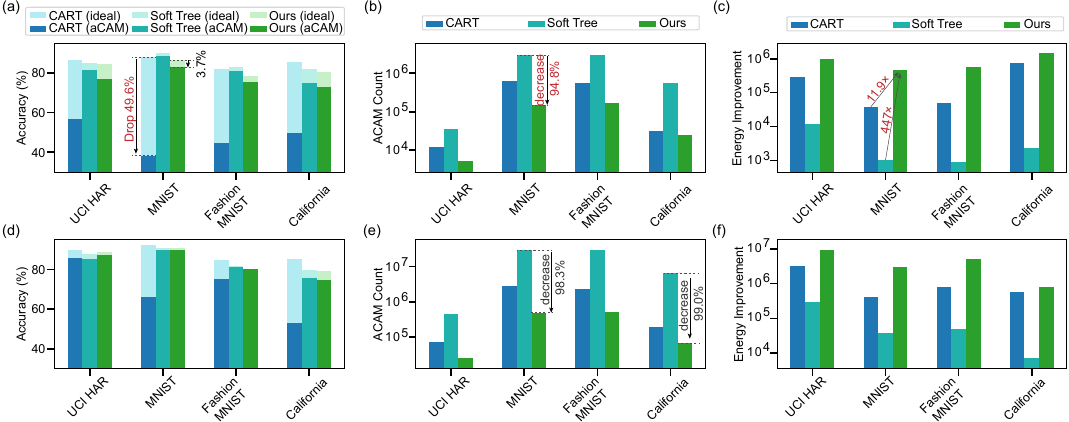}
    \caption{Comparison of ACTIVAR (Ours) against baseline CART and soft-tree models for both single trees (\textbf{a}–\textbf{c}) and ensemble forests (\textbf{d}–\textbf{f}). (\textbf{a}, \textbf{d}) Classification accuracy under ideal software conditions versus hardware-mapped analog CAM arrays. ACTIVAR demonstrates superior robustness to hardware non-idealities. (\textbf{b}, \textbf{e}) Hardware resource usage represented by analog CAM cell count. ACTIVAR’s hierarchical structure reduces resource waste compared to soft tree. (\textbf{c}, \textbf{f}) Energy improvement relative to a general-purpose processor (Apple M2).}
    \vspace{-1em}
    \label{fig:forest_application}
\end{figure*}

\subsection{Classification Performance and Efficiency}

The classification performance, summarized in \autoref{fig:forest_application}, highlights the robustness of ACTIVAR against hardware non-idealities. As shown in \autoref{fig:forest_application}(a), standard CART models mapped directly onto analog CAM arrays are sensitive to analog noise. For instance, on the MNIST dataset, the baseline CART suffers a 49.6\% accuracy drop (from 87.9\% to 38.3\%) due to the device variation and input noise. In contrast, ACTIVAR demonstrates resilience. By employing Variation-Enhanced Training, our compact model mitigates these errors, maintaining hardware accuracy with only 3.7\% drop.

Comparisons with the state-of-the-art analog CAM model (soft tree\cite{wen2025trustworthy}) reveal a distinct trade-off between accuracy and hardware efficiency. Although soft tree achieve high baseline accuracy, they require the entire memory array to be active for every inference, incurring significant energy penalties. Additionally, the structural mismatch between irregular deep trees and rigid analog CAM arrays leads to an exponential waste of cell resources, with soft tree requiring nearly $100\times$ more cells than ACTIVAR (see \autoref{fig:forest_application}(b) and (c)). ACTIVAR overcomes these limitations by utilizing a hierarchical compact tree model that triggers hardware cores sequentially. By selectively executing only the relevant subtrees and avoiding redundant computations in idle branches, ACTIVAR achieves an energy reduction of over three orders of magnitude compared to the full-array activation baseline. 

The advantages of ACTIVAR are most pronounced in ensemble configurations. As depicted in \autoref{fig:forest_application}(d) and (e), the ensemble effect in Random Forests narrows the accuracy gap between ACTIVAR and soft tree. For the MNIST Forest, ACTIVAR achieves an accuracy of 89.8\%, comparable to the soft tree’s 89.5\% on analog CAM hardware, yet it does so with the aforementioned massive reduction in resource usage. Compared to general-purpose processors (Apple M2), ACTIVAR delivers energy improvements ranging from $10^4\times$ to over $ 10^6\times$ (Fashion MNIST), establishing it as an optimal solution for resource-constrained edge intelligence.

\begin{table*}[h]
\centering
\caption{PERFORMANCE METRICS OF TREE REGRESSORS}
\label{tab:TreeRegressionApplication}
\setlength{\tabcolsep}{3pt} 
\begin{tabular}{cc ccc ccc ccc}

\hline
\multicolumn{2}{c}{\textbf{Metrics}} & \multicolumn{3}{c}{\textbf{ACAM RMSE}} & \multicolumn{3}{c}{\textbf{ACAM Cell Count (\(\times10^3\))}} & \multicolumn{3}{c}{\textbf{Energy Improment (\(\times10^5\))}} \\
\hline
\textbf{Dataset} & \textbf{Model Type} & \textbf{CART*} &\textbf{Soft Tree} & \textbf{Ours} & \textbf{CART*}&\textbf{Soft Tree} & \textbf{Ours }& \textbf{CART*}&\textbf{Soft Tree} & \textbf{Ours} \\
\hline
\multirow{2}{*}{California Housing} & Tree & 0.196 &1.153 &\textbf{0.151} & 39.6 ($\times$2.34)& 38.4 ($\times$2.27) & \textbf{16.9 ($\times$1)} & 6.90 ($\times$1) & 6.86 ($\times$0.99) & \textbf{13.57 ($\times$1.97)}
\\ 
 & Forest & 0.153 & 0.544 & \textbf{0.147} & 326.0 ($\times$2.60)& 318.7 ($\times$2.54)& \textbf{125.6 ($\times$1)} & 5.05 ($\times$1)&4.92 ($\times$0.97)& \textbf{8.83 ($\times$1.75)}
\\ 
\multirow{2}{*}{Abalone} & Tree & 0.213 & 1.532 &\textbf{0.094} & 24.2 ($\times$7.88)&23.8 ($\times$7.75)& \textbf{3.1 ($\times$1)} & 12.25 ($\times$1) &12.79 ($\times$1.04) & \textbf{22.48 ($\times$1.84)}
\\ 
 & Forest & 0.106 & 0.372 & \textbf{0.096} & 123.6 ($\times$10.73)& 119.8 ($\times$10.40) &\textbf{11.5 ($\times$1)} & 8.32 ($\times$1)&6.76 ($\times$0.81)& \textbf{10.06 ($\times$1.21)}
\\ 
\multirow{2}{*}{Wine (White)} & Tree & 0.186 &1.607 & \textbf{0.139} & 24.2 ($\times$2.52)& 21.5 ($\times$2.24) & \textbf{9.6 ($\times$1)} & 4.03 ($\times$1)&4.03 ($\times$1)& \textbf{7.81 ($\times$1.94)}
\\ 
& Forest & 0.131 &0.989& \textbf{0.128} & 120.2 ($\times$4.97)& 103.7 ($\times$4.29)&\textbf{24.2 ($\times$1)} & 7.08 ($\times$1)&6.05 ($\times$0.85)&\textbf{9.86 ($\times$1.39)}
\\ 
\multirow{2}{*}{Wine (Red)} & Tree & 0.254& 1.966 & \textbf{0.140} & 13.4 ($\times$7.00)& 9.6 ($\times$5.00)& \textbf{1.9 ($\times$1)} & 19.29 ($\times$1)&19.40 ($\times$1.01)& \textbf{33.65 ($\times$1.74)}
\\ 
 & Forest & 0.143&1.127 & \textbf{0.137} & 44.9 ($\times$2.17)& 34.9 ($\times$1.69)&\textbf{20.7 ($\times$1)} & 12.81 ($\times$1)& 12.26 ($\times$0.96)&\textbf{15.23 ($\times$1.19)}
\\  
\multirow{2}{*}{Appliances Energy} & Tree & 0.172 &0.260 & \textbf{0.089} & 39.9 ($\times$5.20)& 36.5 ($\times$4.75)& \textbf{7.7 ($\times$1)} & 1.81 ($\times$1) &1.77 ($\times$0.98)&\textbf{6.65 ($\times$3.67)}
\\ 
 & Forest & 0.096 &0.130 & \textbf{0.089} & 206.2 ($\times$4.04)& 191.6($\times$3.75) & \textbf{51.1 ($\times$1)} & 3.31 ($\times$1)&2.55 ($\times$0.77)& \textbf{7.63 ($\times$2.31)}
\\ \hline
\multicolumn{11}{l}{\footnotesize *: Classification and Regression Trees (CART*)\cite{breiman2017classification}, implemented by scikit-learn} \\
\end{tabular}
\end{table*}
\subsection{Regression Performance and Efficiency}
We extend this evaluation to regression tasks, measuring performance via Root Mean Square Error. As detailed in \autoref{tab:TreeRegressionApplication}, ACTIVAR regressors consistently achieve lower (better) RMSE values compared to baseline (CART and soft tree) implementations under non-ideal hardware conditions. Beyond predictive stability, the proposed compact architecture  lowers hardware overhead, delivering a 
$1.2\times$ to $7.9\times$ reduction in analog CAM cell utilization compared to standard regression trees realized on analog CAM. This further translates to a $1.7\times$ to $3.7\times$ reduction in energy consumption across the tested datasets, validating that the performance and efficiency benefits of ACTIVAR are preserved in regression domains.
\section{Conclusion}
\label{sec:conclusion}

In this paper, we propose ACTIVAR, a hardware-software co-designed framework for accelerating decision tree inference using analog Content-Addressable Memory. We address the fundamental structural mismatch between algorithmic tree models and analog hardware arrays by introducing an analog CAM-aware compact tree model, which reduces required computations by 99\% compared to state-of-the-art models when mapped to analog CAM arrays. To overcome hardware non-idealities such as noise and device variation, we propose Variation-Enhanced Training and a novel hard-boundary cell design. Evaluated on MNIST, ACTIVAR limits accuracy degradation to just 3.7\% under realistic analog noise, whereas baseline models experience a 49.6\% drop. Benchmarking results demonstrate that our pipelined multi-core architecture achieves energy efficiency gains of $10^4$ to $10^6$ times over commercial CPU. By co-optimizing across the model structure, training strategy, and circuit levels, ACTIVAR enables scalable, resilient, and high-capacity machine learning inference for resource-constrained edge devices.

\bibliographystyle{IEEEtran}
\bibliography{references}
\end{document}